# Effect of Randomly Distributed Asymmetric Stone-Wales Defect on Electronic and Transport Properties of Armchair Graphene Nanoribbon

Mobin Shakeri[1]

*Department of Physics, Kharazmi University, Mofateh Ave., Tehran, Iran.*

**Abstract :** By applying tight binding model, we investigate the electronic and transport properties of randomly distributed Stone-Wales (SW) defects on an armchair graphene nanoribbon (AGNR). We use four different functions, as distribution functions, to generate our SW defected nanoribbons. It is found that defect density can have a major effect on the conductance of our defected system, whilst other configurations such as defect orientation will contribute less. In our investigations, some special geometries are found which shows interesting electronic and transport properties. These special cases along with the other data provided can be used to engineer band gap, electronic properties and transport properties of graphene nanoribbons to meet a desired purpose.

## I. Introduction

Graphene is a single layer allotrope of carbon arranged in a hexagonal lattice, which attracted a great deal of attention ever since its fabrication by Nobel Laureate KS Novoselov, et al. in 2004 [1]. It has some extraordinary thermal, electrical and mechanical properties, however, amongst these, the unique electronic properties are considered to be the most fascinating aspect of it. Graphene is a material with zero band gap, and the main difficulty regarding the practical applications of it has been this lack of band gap in it [2]. To extend graphene's application in electronic devices, it is important to open and tune the band gap of it, without changing its' exceptional properties [3]. Many experimental [3] and different fabrication strategies [4], including epitaxial growth [5], substitutional doping [6], quantum confinement [7], and chemical functionalization [8-9] have been proposed. Creating a narrow quasi 1-dimensional (1D) nanoribbon from a 2-dimensional (2D) graphene sheet can be useful for creating band gaps and other applications in electronic devices [10-11].

Graphene nanoribbon (GNR) is a strip of Graphene, which could be tailored for many purposes in the field of quantum transport. Classification of GNRs divides them into two main groups based on their edges: zigzag graphene nanoribbons (ZGNRs) and armchair graphene nanoribbons (AGNRs). It has been noted that degenerate edge states of ZGNRs make them a quasi-metallic material, however, AGNRs can either be metallic or semi-conducting, based on their width [12-13].

GNRs show excellent electronic properties when they are formed in an outstanding perfect atomic lattice, however structural defects, which may form during growth or processing, worsen the performance of graphene-based devices [14]. However, varying the system from its perfect configuration can be useful in some applications, since it will be possible to alter the local properties of graphene and to obtain new functionalities [15-16]. In addition to these mentioned defects which are structural and hard to avoid [14], defects can be intentionally introduced into graphene, by numerous techniques such as irradiation or chemical treatments [17]. Graphene defects can be seen in different types, such as (1) stone-wales(SW) [18-19], (2) single vacancy (SV) [20-21], (3) double and multiple vacancy (DV) [22] , (4) line vacancy [23]; it is possible to generate them through different mechanism and techniques such as (1) crystal growth [24-25], (2) irradiation with energetic particles [26-27], (3) chemical treatment [28]; and many healing procedures and methods can be used such as (1) Healing by absorption [29], (2) Self-healing [30], etc. [14]

Modern electron microscopes in recent experimental research, include aberration-corrected condensers, which will enable us to focus an electron beam upon a very tiny spot of approximately 1 Å in diameter, thereby forming vacancies with around atomic selectivity [31]. Because of this, an exact modification of the material's atomic structure is now possible through introducing point defects [32].

SW defect, which involves the change of connectivity of two π-bonded carbon atoms, leading to their rotation by 90° with respect to the midpoint of their bond, possess the lowest transformation energy amongst all defects in graphene-based systems. Also due to its low formation energy of 5 eV, it has been confirmed to be energetically more favorable than others [14, 33].

---

[1]:Mobin.shakeri@live.com

There have been many kinds of research regarding the transport properties of graphene's different forms, under the influence of different types of defect presence [34, 13, 16, and 23]. Different graphene defect types and their corresponding transmission characteristics has been explored by Sheng Chang et al. [35]. Effects of different SW defect Symmetries has been studied by Jun Zhao et al., on the electronic structure and transport properties of narrow AGNR [36], after it was argued that AGNRs are energetically more favorable, compared to ZGNRs, and that the former has been more often described in experimental observations [37]. In these studies, it has been revealed that the presence of an SW defect could be favorable for electron transport and it has been found that the presence of asymmetric SW could enhance electron transport by 13%.

Even though the importance of studying random defect's distribution is realized, only some scattered studies have been executed upon these types of effects. J. P. C. Baldwin et al., have studied "the effect of random edge-vacancy disorder in zigzag graphene nanoribbons" in 2016 [38]. A study of "Rupture of graphene sheets with randomly distributed defects" was also done in the same year [39]. Other studies on zigzag nano ribbon are: [50, 51]. Studies around the SW defects were mainly around nanotube structure. Most recently, a study was carried out on Zigzag Graphene Nanoribbons with Stone-Wales Defect [52]. The most important random SW defect distribution study was done by Qiang Lu and Baidurya Bhattacharya, on the mechanical properties of carbon nanotubes [40].

In this paper, we have reported the effects of randomly distributed Asymmetry SW defects on the transport properties of AGNRs. This novel configuration with the current statistical method is a follow-up of the previous papers. An SW defect is selected because of its' low formation energy and high simplicity of the structure, which doesn't include any added or removed atoms [14, 33]. AGNR is selected over ZGNR, due to its energetical stability [37], and an Asymmetric case of SW defects has been selected, because of its promising electron transport enhancement as shown in [36]. We have calculated the band gap and conductance of 4 different distributions, under different defect densities and two different orientations of SW defects using these properties, in order to find the most suitable placement and geometry of SW defects in a GNR. The models and methods of this study are discussed below.

## II. Model and Computational Methods

In this research, we have used AGNRs over ZGNRs due to the fact that the former could be energetically more stable, compared to the latter [37]. AGNRs are divided into three classes, due to their width [41-42]: (1) $N = 3m + 2$; (2) $N = 3m + 1$; (3) $N = 3m$, where m is an integer number. The periodic AGNRs with $N = 3m + 2$ shows metallic behaviour, whereas the other two show semiconducting behaviour within the tight binding (TB) formalism. We have selected N=14 for our simulations, to check the band gap opening of different defect distributions. As it was discussed above, among the different defects, an SW defect is selected because of its' low formation energy [14, 33, 40].

We have selected asymmetric SW (ASW) distribution due to Jun Zhao et al., research, which indicated that the presence of asymmetric SW could enhance electron transport by 13% [36].

We have used the DFT open source package for material eXplorer (OPENMX3.8) [43], to relax the nanostructure of single ASW, in the middle of N=14 AGNR using a supercell presented in Fig.1. We have used Local-density approximation (LDA) as an exchange-correlation potential for our self-consistent field (SCF) calculations. The Brillouin zone was sampled by a 200 * 10 * 1 Monkhorst-Pack k-point mesh [44]. The SCF energy criterion was set to 1.0e-6 Hartree. The geometry optimization was done by the steepest descent (SD) method with a maximum iteration number of 20, until the maximum force on atom has become smaller than 1.0e-4 (a.u.). The result is shown in Fig. 1.

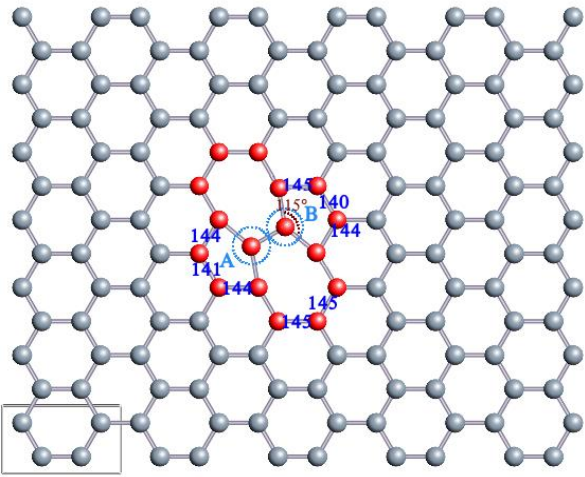

**Fig**. 1: Optimized structure of a single stone-wales defected graphene nanoribbon with the width of N=14 atoms. The grey atoms represent carbon atoms. The Highlighted red atoms show the defect atoms. The squared area represents the unit cell used in our calculations.

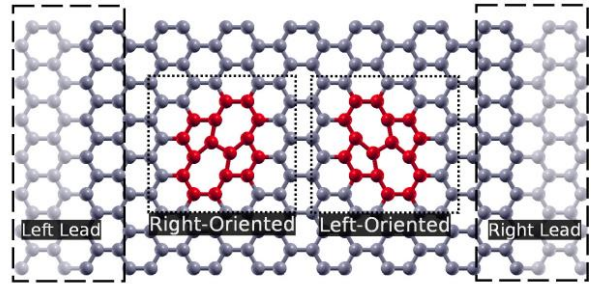

**Fig**. 2: Schematic view of right-oriented and left-oriented asymmetry stone-wales defect. As it is shown, the right and left are based on the position of leads.

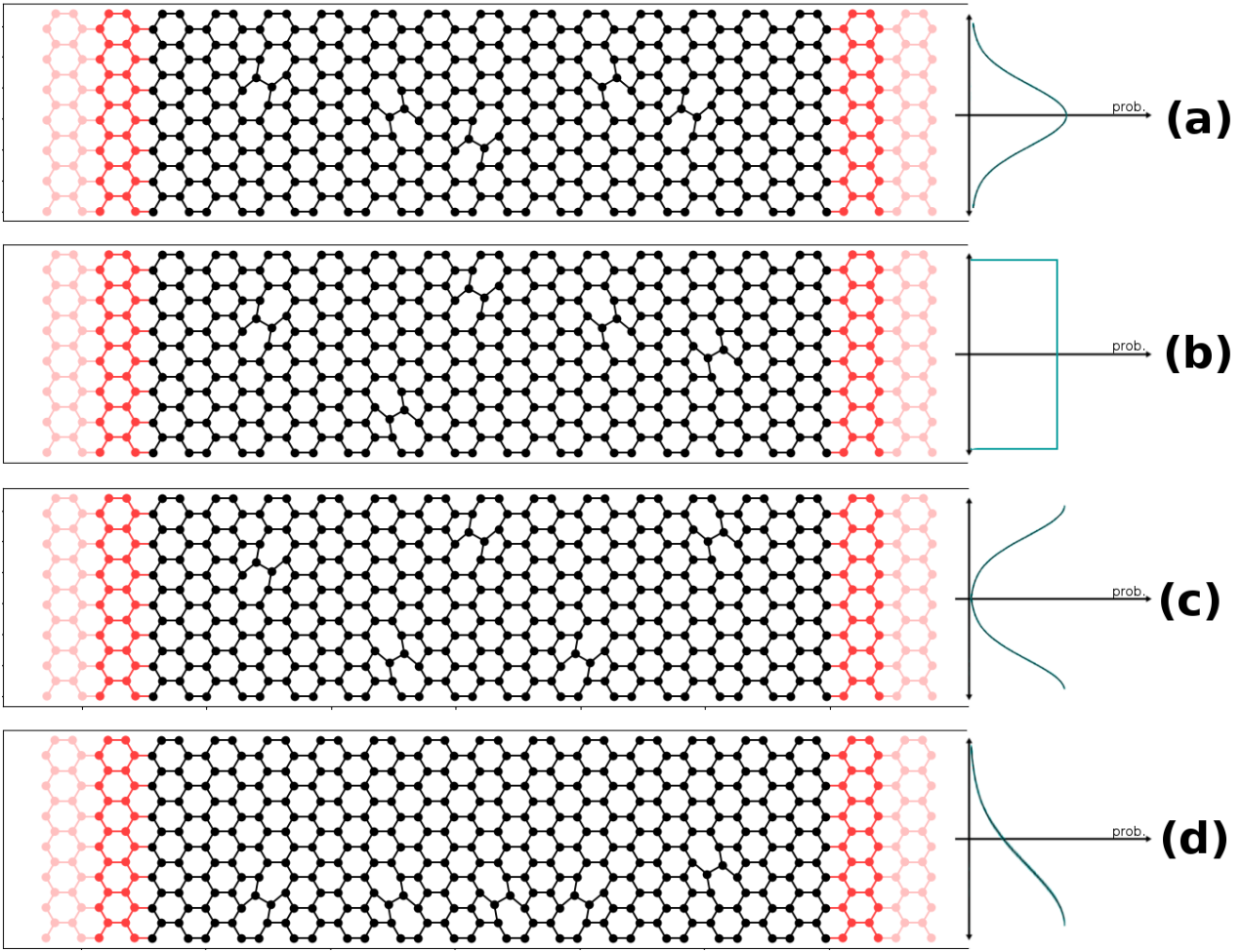

**Fig**. 3: Representation of the four distribution functions along with an example generated on a graphene nanoribbon. the selected names for each function is (a) Gaussian Distribution (b) constant distribution (c) two-edge Gaussian Distribution and (d) one-edge Gaussian Distribution. the red areas represent the leads.

We have used python programing language to program a weighted random defect generator using four longitudinal distribution functions along the width of our AGNR. the functions are (1) constant weight, (2) Gaussian normal distribution at the centre of AGNR, usually referred to "Gaussian distribution" in this paper, (3) Gaussian normal distribution at the edges of AGNR, considering a periodic boundary condition, referred as "Two-Edge Distribution", and (4) one-edge Gaussian normal distribution which vanishes at the other edge and is usually called "One-Edge Distribution" in this paper. The functions and an example of their generated system are shown in Fig.3. These particular functions were chosen based on the previous researches, which have shown the defects generated around the edges might have less negative consequences on the transport properties of GNRs [16, 38]. The fourth function was chosen due to the interesting results that were seen from the first three ones, which will be discussed later.

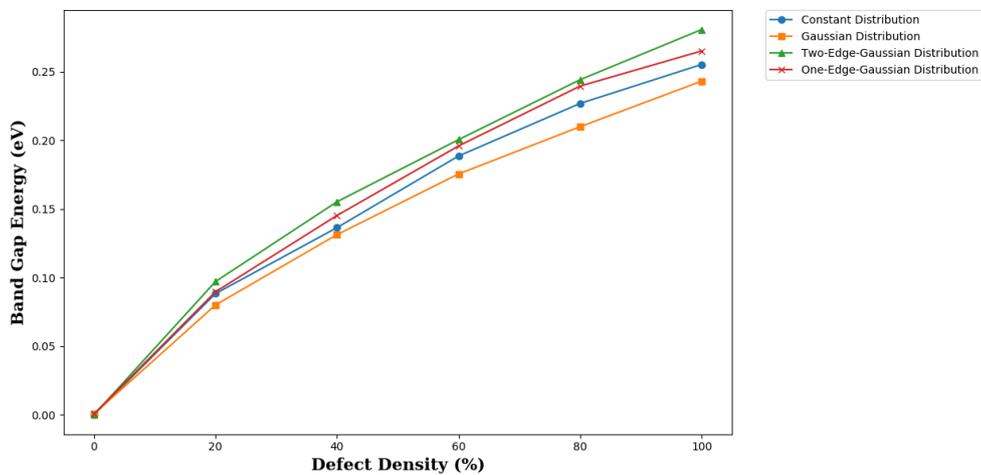

**Fig**.4: Band gap energy of four different distributions, as a function of defect density of the scattering region.

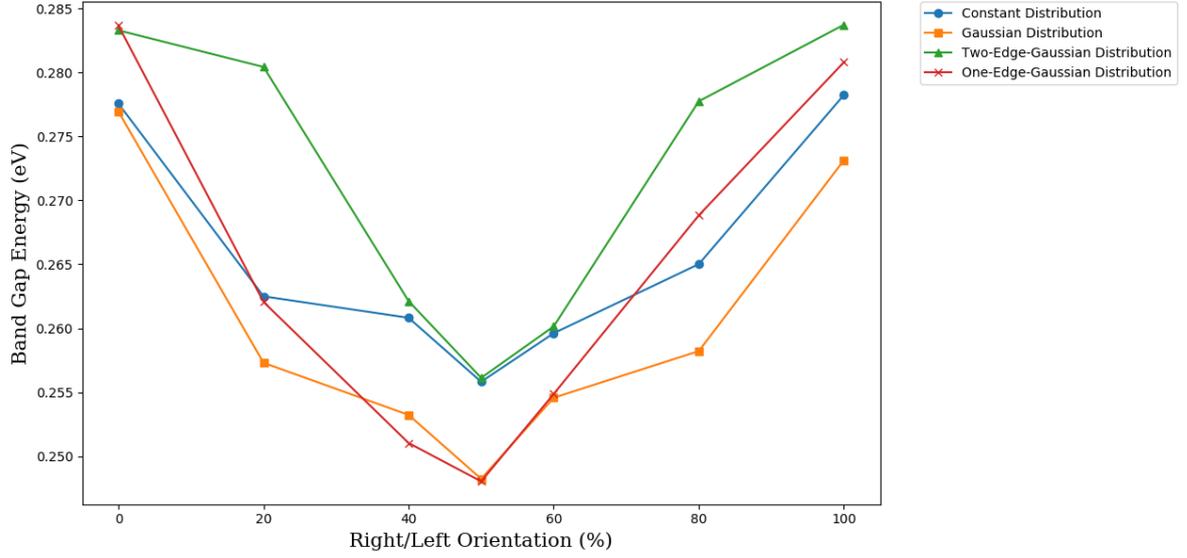

**Fig**. 5: Band gap energy of four different distributions, as a function of right/left oriented defects ratio on the scattering region.

It should be noted that the ASWs, can occur in two different right-oriented and left-oriented shapes as shown in Fig. 2. A variable which indicates the ratio of the number of right-oriented to the number of left-oriented ASWs is taken into account.

We have assumed that the relaxed structure of a single ASW is the same as the structure of a longitudinal distribution of them. The generated weighted random distribution of defects can also be affected by the density of defects. Because of the narrow selection of our AGNR, placement of multiple ASW along the width of our ribbon is not expected, therefore a defect density is defined, only along the length of our ribbon. This density is defined as the ratio of the number of defects on longitudinal 2 * 7 supercell. This special selection is due to a distinction of defects on nanoribbon, similar to pristine methods that have been done earlier [40]. The unit cell is presented in Fig.1.

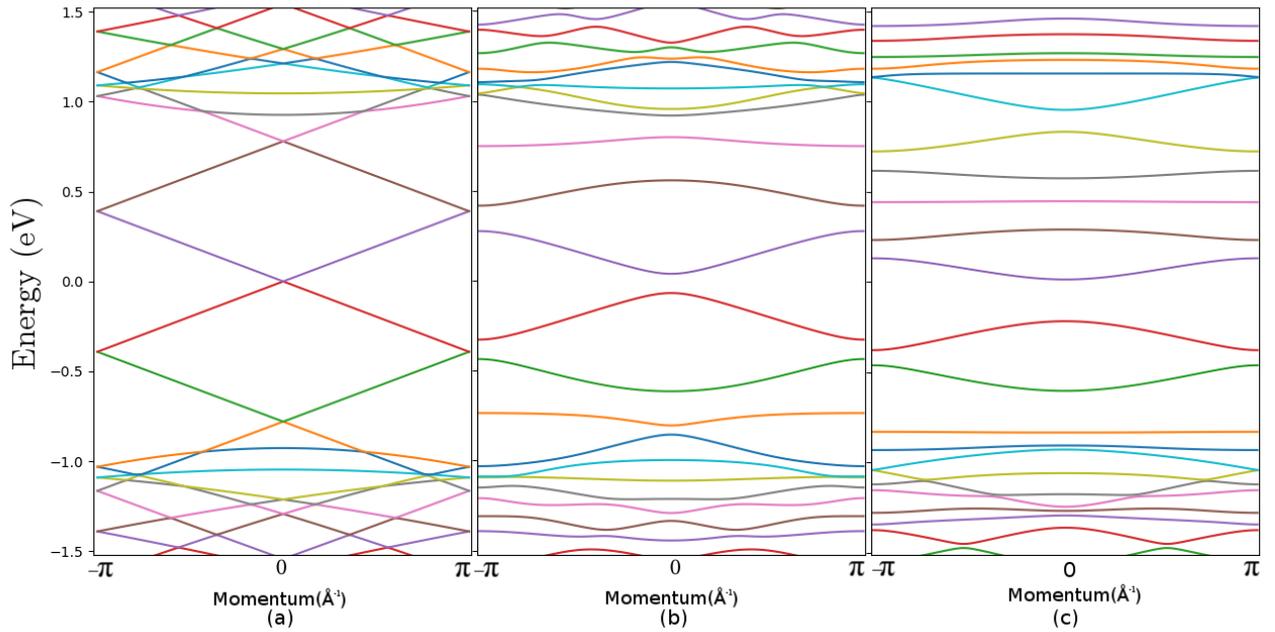

**Fig**. 6: Band structure of three ASW defected AGNRs with different defect densities where in (a) the defect density is 0%, in (b) it is set to 20% and it is 40% in (c). The Fermi level is set to 0 eV.

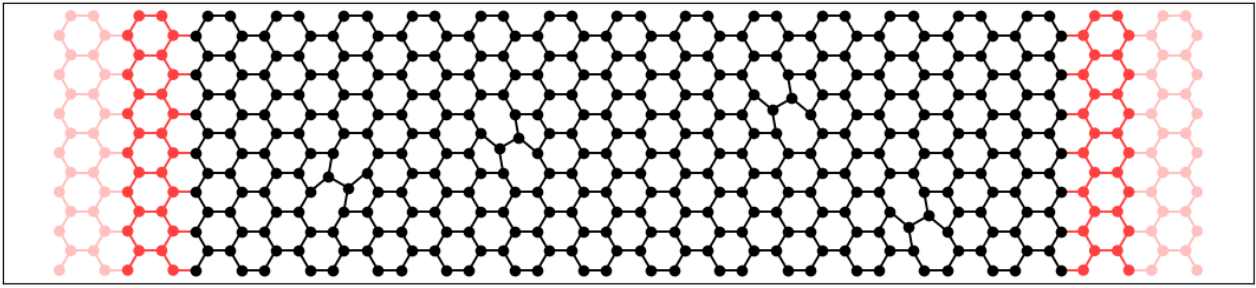

**Fig**. 7: An example of randomly generated Asymmetric stone-wales defected Armchair Graphene nanoribbon using our python defect generator code. All of these AGNRs are generated based on the four defined distribution functions. The black atoms represent carbon atoms, and the red ones represent the leads.

After the defected systems were generated, we have used them on Kwant [45], a software package for quantum transport, to create a tight binding model out of them. The third neighbour approximation was used for these systems with the on-site energy of 0 eV and first neighbour hopping of 2.74 eV. The rest of the energies are mentioned in [46]. Using these settings, gap energies and conductance spectra were calculated by Kwant package. Most of the results were taken from the average behaviour of multiple randomly generated systems. Calculations in conductance spectra that were used only for a qualitative comparison were done with a first-neighbour approximation, because of the resulting higher speed of calculations and that their relative accuracy is sufficient for our purpose.

A certain criteria is chosen for the convergence of the results, when the calculation is based on many randomly generated systems. Based on these criteria, The Error for this average is around 1%. Another Criterion is chosen for the orientation of ASWs, and this Criterion of convergence is for the symmetry seen in Fig.5.

A further discussion of these results is given in the next section.

## III. Results & Discussion

First, a band gap opening was investigated under different circumstances of ASW defected AGNRs. Fig.4 and Fig.5 represent the results. As it can be seen from Fig.4, the bandgap width expands as the defect density is increased, and this behavior is visible for all of the four different distributions. It should be noted that, based on Fig.4, a Gaussian distribution of ASWs around the middle of AGNR increases less energy gap, as the number of ASWs increases. It should be noted that this relation is not linear; meaning for each new ASW, less amount of energy gap is broadened. As an example, for a constant distribution of ASWs, one ASW creates a band gap of around 0.08ev, while the band gap for the fifth added ASW was only broadened by 0.02 eV. It can also be inferred that Fig.4 shows less band gap broadening for distributions concentrated around the middle of Graphene nanoribbon.

Energy gap can also be affected by the ratio of right/left oriented ASWs. The result is represented in Fig.5. As it can be seen, a less coherent result is obtained, compared to Fig.4. A point at 60% right/left ratio can be seen as a meeting point of all distributions. This point also acts a minimum for one and two-edge Gaussian distribution. We can infer that for these distributions, the band gap broadens as the ASWs become uniformly right or left oriented. But the same is not true for the other distributions. The energy gap difference order is also much lower than Fig4, which means that the orientation of ASWs has less effect on energy gap opening, compared to ASW defect density.

Fig.6 shows the band dispersion for 3 different states, with the same distribution and right/left orientation ratio, but different defect densities. Even though the standard band structure is not suitable for this highly disordered materials, this spectra is more useful for band gap calculations than Kwant KPM DOS calculations, due to its high numeric errors near the 0 energy [45]. These diagrams are represented with the nearest-neighbor approximation only. It can be seen that a band gap opens near the Fermi level. It can be seen that the bands near the Fermi level have lifted and deformed after the ASWs are introduced. Valence and conduction bands change to a dispersionless tail band extending to $\pi$ point. By adding more ASWs, more energy gap is visible, along with more bands becoming a dispersionless tail band on higher energy levels. This is only an example of the general behavior of band dispersions, due to the increment of defect density. These results are generally consistent with previous reports [47, 48, 36], and the inconsistencies are due to the Tight-Binding model that is used here, compared to previous density functional theory (DFT) works. These differences were compared and discussed before [49].
After the calculation of band dispersion, transport properties of these ASW defected AGNRs are going to be discussed. Fig.7 represents one example of defected AGNR, which is connected to two AGNR leads with the same width. All of our generated systems with ASW defects are placed in the same configuration. All of the conductance calculations were done from the left lead to the right lead.
Since getting the average for transport properties does not show us any meaningful results, another approach is preferred at this step. For each different distribution, a most probable configuration is selected, and the results are reported according to the ASW defect generated system.
Fig.8 shows the conductance spectra of different distributions with the same defect density and the right/left orientation ratio. The length of this scattering region is set to be 46 atoms long, consisting of a maximum of 5 defects along it. From this result, we can see that only Gaussian distribution has a non-zero value around the Fermi energy, which is set to zero. The general behavior of these distributions is the same at higher energy levels. We can see that the constant distribution generates distant peaks at lower energies. Most of the low energy level peaks are placed around ±0.5 eV which is due to the energy gap broadened caused by the existence of ASW defects. The seen valleys occur due to the backscattering of the electrons from the localized states on the ASW defect atoms.
Since the low energy peaks and valleys are sensitive to small changes of defect placement and orientations, we cannot infer more information from these spectra, and a further discussion is needed for special cases to find a suitable pattern amongst them. The result of the conductance of AGNRs with the different ASW's right to left orientation ratio is presented in Fig.9, within 4 different distributions with a fixed defect density. As it was expected, the conductance of the scattering region, which is defined from the left lead to the right, behaves generally the same at the higher energies, with respect to Fermi level, under the variation of orientations.

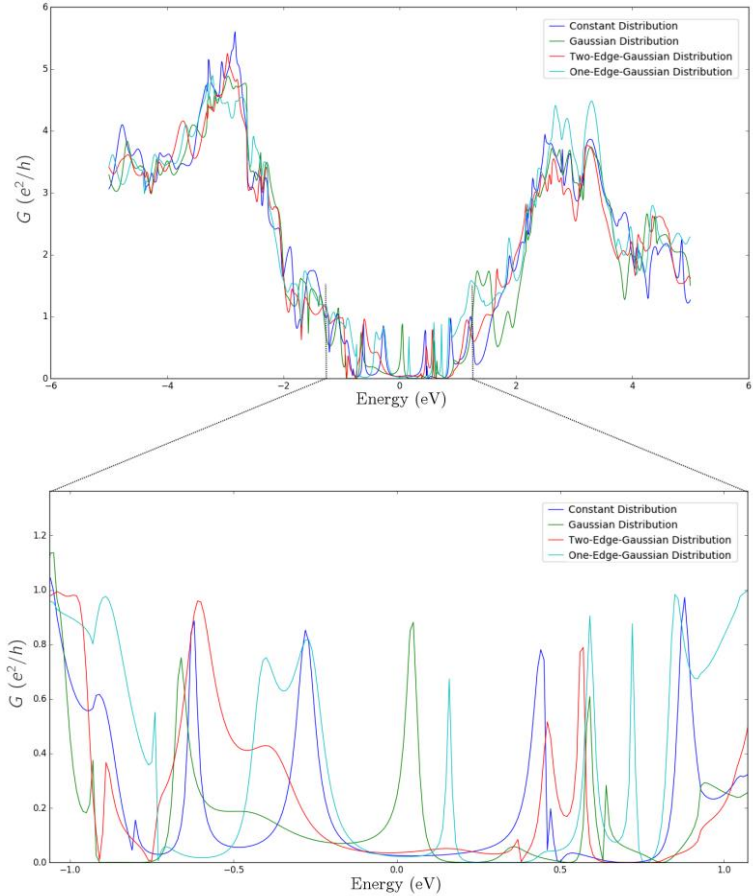

**Fig**.8: Conductance spectra of the four distributions as a function of electron energy at the same conditions such as defect density and right/left oriented ratio. The diagram below shows a detailed magnified spectrum of the lower energy levels.

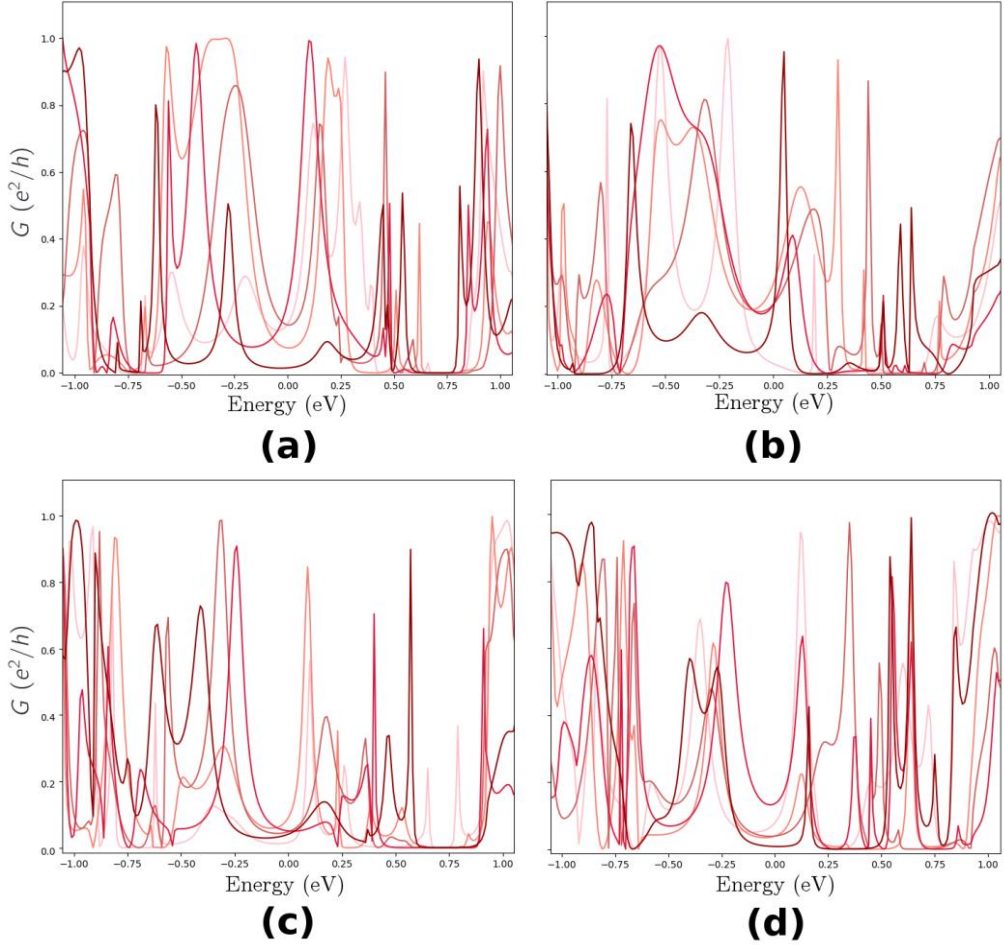

**Fig**. 9: Conductance spectra of the four distributions as a function of electron energy with a different right/left orientation ratio. (a), (b), (c) and (d) represent a constant distribution, Gaussian distribution, two-edge Gaussian distribution and one-edge Gaussian distribution respectively.

Comparing the cases with all of ASWs oriented to the right or to the left shows some interesting results. As it can be seen from Fig.9 (b), conductance peak at 0 eV that was discussed earlier, is turned to almost zero, when all of the orientations are switched. The new peak is now located at around -0.2 eV.

It can be seen from the diagrams in Fig.9 that the distributions which concentrate ASWs at one edge, behave more similarly with the different orientation ratio, than distributions like constant or Gaussian, which spread ASWs along AGNRs' width. This could mean neighboring ASWs with a different orientation, which are placed diagonally (placed at different length and width, compared to length and width of the nanoribbon) can affect the total transport properties of nanoribbon, way more than two ASWs which are placed at the same width. A further research can be carried out regarding these phenomena.

Fig.10 shows the resulting conductance of the scattering region, under various defect densities, for different ASW distributions while all the ASWs are right-oriented. It is visible from the different results of Fig.10, that the conductance spectra are sensitive to defect density. It can be seen that conductance at the higher energies of all distributions, is heavily affected by the density of defects in our scattering region, and denser defects result in an average lower conductance. By comparing Fig.10 to Fig.9 and Fig.8 we can easily see a visible general pattern in the former, despite the two latter. This could mean that not only the varying defect densities have the most meaningful effects, but also this property, has the largest effect on the transport properties of ASW defected GNRs. Since the highest defect densities in Fig.10, averagely have zero conductance in low energy ranges, we can infer that the higher defect density results in a worsened transport property of AGNR. However, these high-density generated AGNRs possess peaks, usually as high as the low density systems in the energies near Fermi level. These results were expected from Fig.4 and Fig.6, since these added ASWs, broadens the gap energy, and creates dispersionless tail bands near the Fermi level. This results in zero conductance channels near the Fermi level. We can see from Fig.10 (b) that the

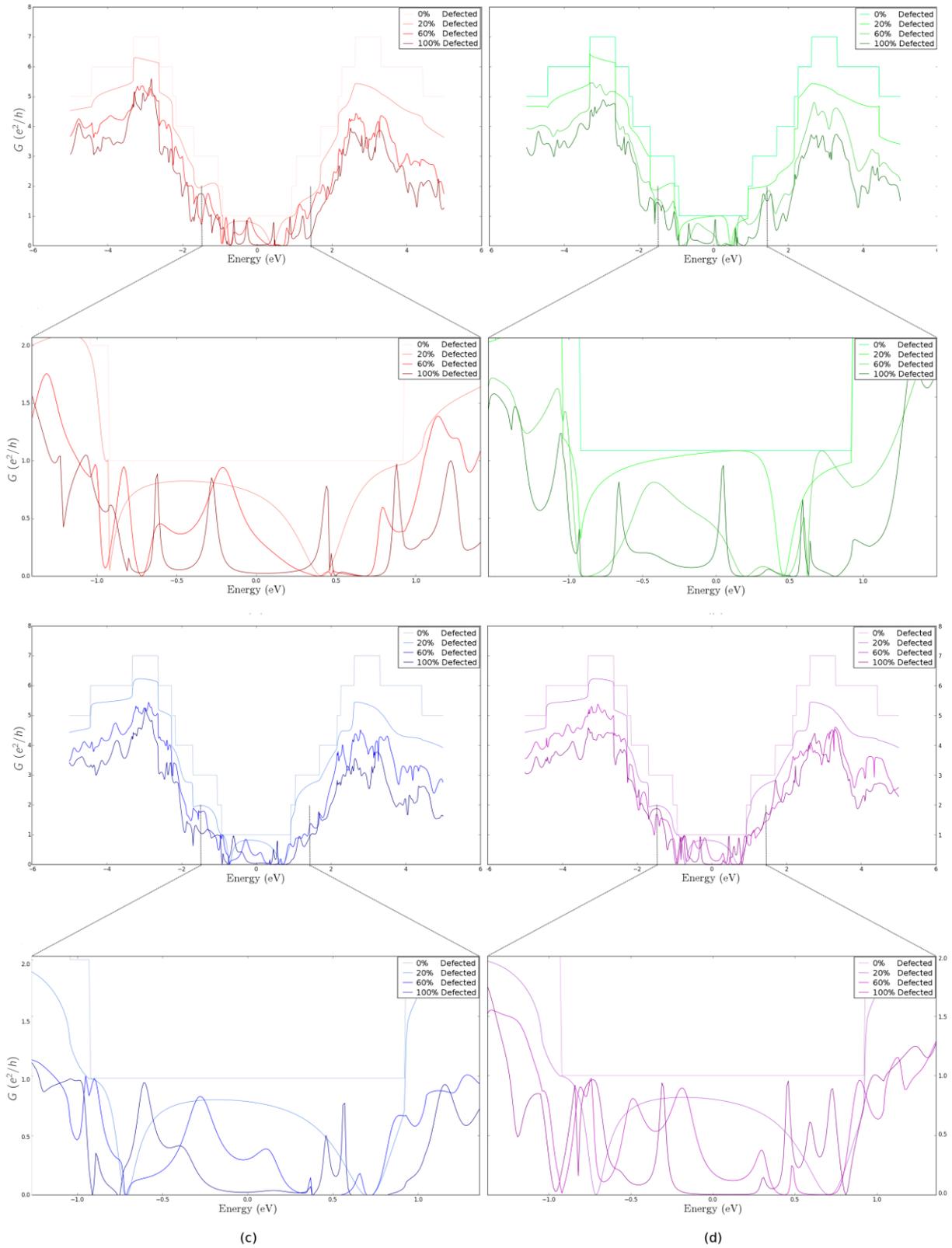

**Fig**. 10: Conductance spectra of the four distributions as a function of electron energy with different defect density. The magnified low-level energy spectra are shown below. (a), (b), (c) and (d) represent a constant distribution, Gaussian distribution, two-edge Gaussian distribution and one-edge Gaussian distribution respectively.

Gaussian distribution at the highest ASW defect density has a non-zero conductance at the Fermi level. In these randomly generated systems, some interesting configurations were encountered with special transport properties. In the simplest cases of these special configurations, multiple right-oriented

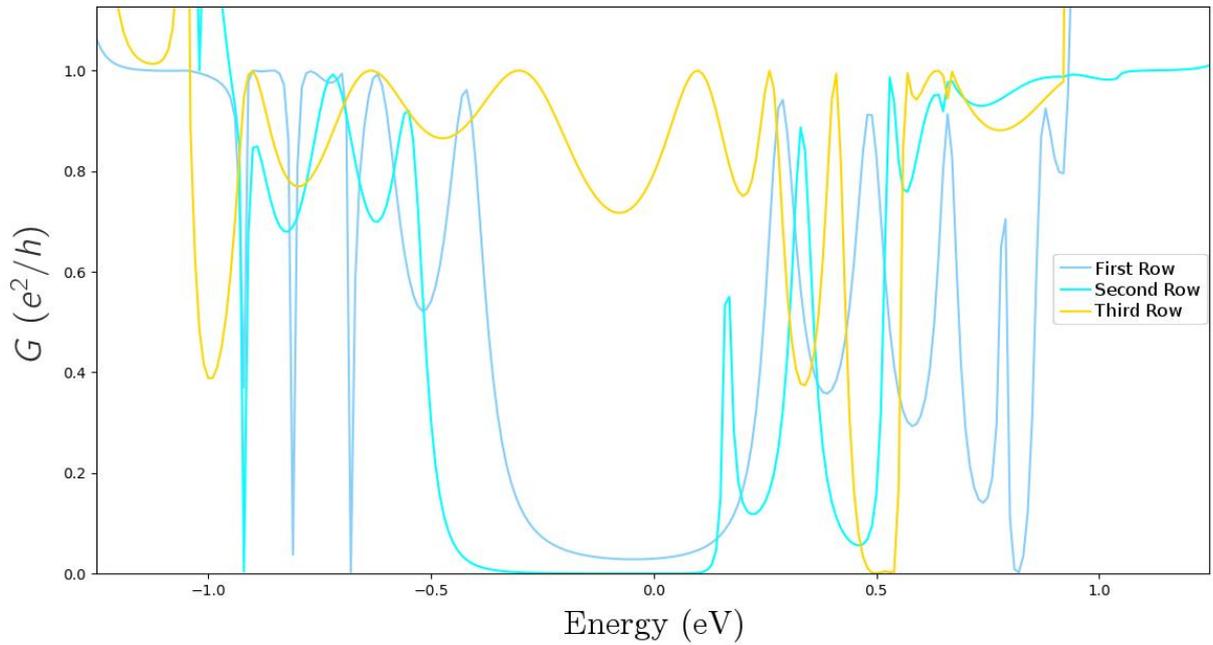

**Fig**. 11: Conductance spectra of three special configurations of ASWs in AGNR as a function of electron energy. These special systems consist of 5 ASW defect in the first, second and third row of AGNR, and is labelled as First Row, Second Row, and third row respectively.

ASWs can be put into one straight line at the same width along the AGNR. This line can be placed in three places, from the very edge, to the center of the nanoribbon. The results are presented in Fig.11. As it can be seen, the ASWs that are placed next to each other in the middle of the GNR acts differently from the other two. This configuration, unlike others, only consists of one valley around 0.5 eV. This can be inferred as if this configuration acts somehow like a single ASW, which is placed in the middle of nanoribbon. It should be noted that just like the Gaussian distribution which is centered on the middle of GNR and has a non-zero value at the Fermi level, this special case is also non-zero at the same energy.

Fig.12 is a comparison of the mentioned special case which is consisting of five ASWs in one row, to the cases with three and one ASWs along the same line.

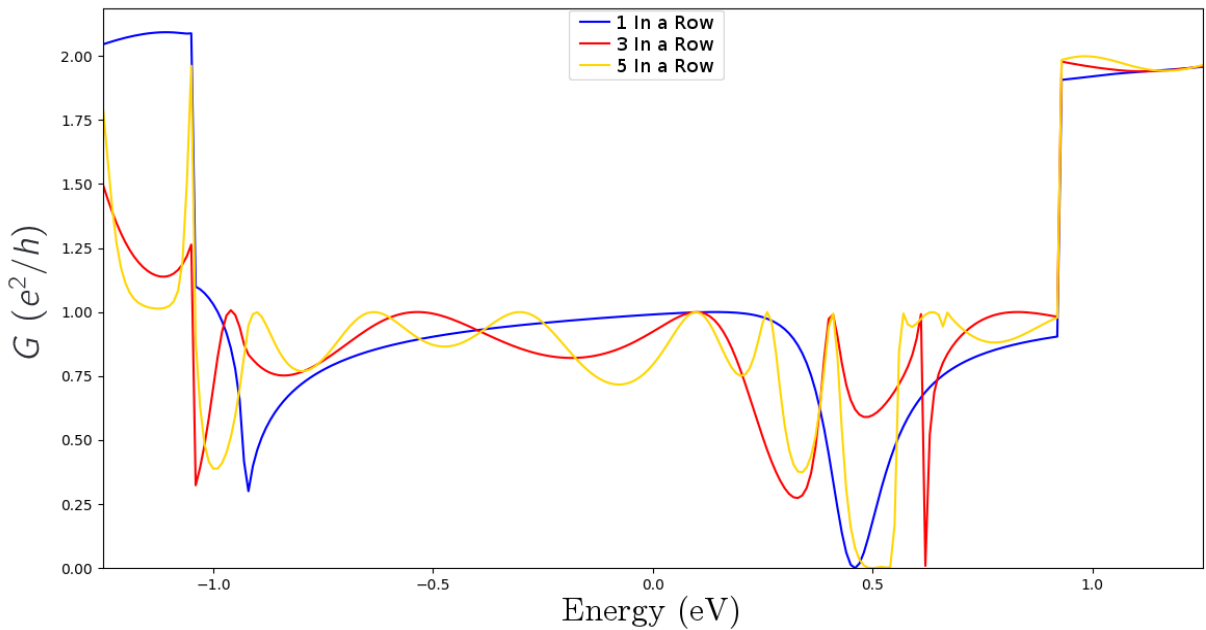

**Fig**. 12: Conductance spectra of three special configurations of ASWs in AGNR as a function of electron energy. These special systems consist of 1, 3 and 5 ASW defects in the middle (third) row of AGNR and are labelled as 1 in a Row, 3 in a Row and 5 in a Row respectively.

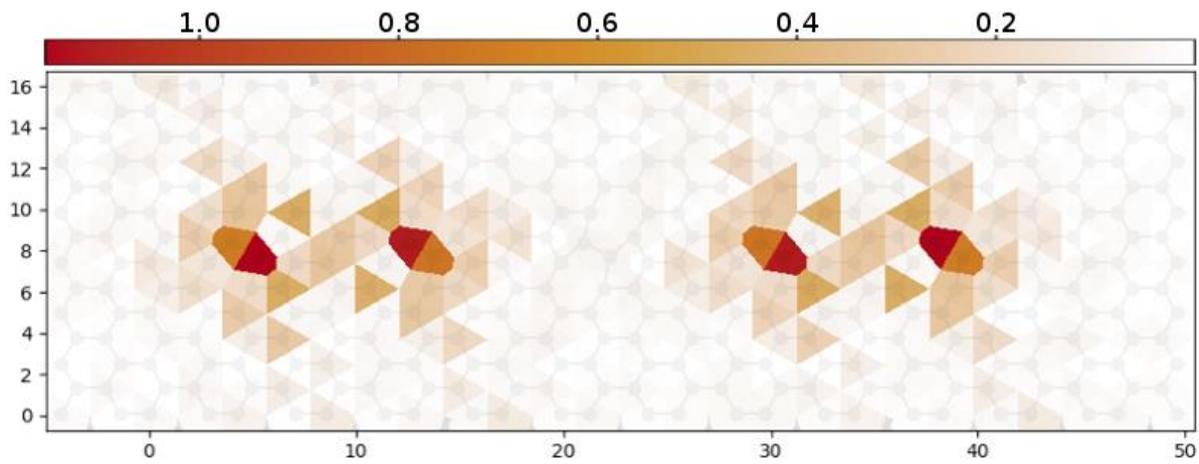

**Fig**. 13: Local Density of States (LDOS) projected on the position space of our ASW defected AGNR. The scale of AGNR position space is shown in Angstrom.

As it can be seen, this set of 5 ASW placed in a special configuration, generally behaves like a single ASW, with a valley around 0.5 eV, and non-zero conductance at the zero energy level.

The zero conductance at exactly 0.5 eV is the result of the backscattering of all, except the middle ASW, which is confirmed via Fig.13. This exact result was previously seen in [37]. This diagram is a local density of state (LDOS) at the same energy, projected into position space of the graphene nanoribbon. This phenomenon is unique to this and higher density cases, and it seems that the middle ASWs that are sandwiched between other ASWs has no role in backscattering of electrons in general. This could be the cause of single conductance valley seen in this special case.

Furthermore, it can be noticed that atoms (A) and (B) (as it is shown in Fig.1) of ASW plays a critical role in this phenomena, as it's evident in Fig.13. It can also be seen that DOS concentrate alternately on atom (A) and the atom (B), in every other two neighboring ASWs in this nanoribbon. A further investigation is needed regarding this phenomena.

These results revealed some more interesting cases of ASW defected AGNR. These special transport behaviors could result in the betterment of the graphene nanoribbon bandgap engineering. However, a further research is needed to continue describing these special cases in a more profound way.

## IV. Conclusion

We have reported a theoretical investigation on randomly distributed asymmetry Stone-Wales defects on armchair graphene nanoribbons, to study their role on the band structures and electron transport properties. It is found that amongst the different degrees of freedom to generate an AGNR, defect density has the most important role on the band and conductance of our defected system. Other data on conductance spectra of different configurations of our 4 different distribution functions are presented in this article. In the end, some special cases were examined, like multiple ASWs placed along the AGNR length in a straight line, and some new properties were found amongst these special configurations, e.g., it was found that this special geometry acts like a single ASW in its' conductance spectra. The results of this article can be used to engineer the bandgap of GNRs and reach a desired system with a desired electron transport property. However, a further investigation is needed in several cases that were discussed, to further expand the subject.

## V. Acknowledgment

We would like to thank A. Vahedi, F. Kanjouri and M. Lashgari for their help and their fruitful comments. We are also grateful to A. Shakeri for reading the manuscript and helpful comments.

The author also thank the referees for their positive comment to improve the quality of this paper.